\documentclass[twocolumn,showpacs,preprintnumbers,a4,prb]{revtex4}
\usepackage{graphicx}
\usepackage{dcolumn}
\usepackage{amsmath}
\usepackage{amssymb}
\usepackage{bm}

\begin{document}

\title{Effective ferromagnetic coupling between a superconductor and a ferromagnet in LaCaMnO/Nb hybrids}
\author{D. Stamopoulos,\footnote[3]{Author to whom correspondence
should be addressed (densta@ims.demokritos.gr)} N. Moutis, M.
Pissas and D. Niarchos}

\affiliation{Institute of Materials Science, NCSR "Demokritos",
153-10, Aghia Paraskevi, Athens, Greece.}
\date{\today}

\begin{abstract}

In this work we present magnetization data on hybrids consisting
of multilayers (MLs) of manganites
[La$_{0.33}$Ca$_{0.67}$MnO$_{3}$/La$_{0.60}$Ca$_{0.40}$MnO$_{3}]_{15}$
in contact with a low-T$_c$ Nb superconductor (SC). Although a
pure SC should behave {\it diamagnetically} in respect to the
external magnetic field in our ML-SC hybrids we observed that the
magnetization of the SC follows that of the ML. Our intriguing
experimental results show that the SC below its $T_{c}^{SC}$
becomes {\it ferromagnetically} coupled to the ML. As a result in
the regime where diamagnetic behaviour of the SC was expected its
bulk magnetization switches only whenever the coercive field of
the ML is exceeded. By employing specific experiments where the ML
was selectively biased or not we demonstrate that the ML inflicts
its magnetic properties on the whole hybrid. Possible explanations
are discussed in connection to recent theoretical proposals and
experimental findings that were obtained in relative hybrids.

\end{abstract}


\maketitle

In recent years hybrids comprised of superconducting (SC) and
ferromagnetic (FM) ingredients have attracted much interest due to
the fascinating properties that they exhibit depending on the
choice of materials, their specific structure etc. Among others,
basic structures of such artificial hybrids are bilayers (BLs),
trilayers (TLs) and even multilayers (MLs) of homogenous films but
also distinct ferromagnetic nanoparticles (FNs) placed on top of
or embedded in a SC
layer.\cite{Otani93,Ketterson98,Martin99,Lange03,Stamopoulos04SST,Stamopoulos05PRB}
For the case of ordered arrays of FNs it has been shown that the
magnetoresistance of the SC presents distinct periodic minima at
specific values of the applied magnetic
field.\cite{Otani93,Ketterson98,Martin99} When the FNs are
randomly distributed in the SC, periodic minima are not observed
in the magnetoresistance but a clear enhancement in the
surface-superconductivity critical field occurs when their
saturation field is
exceeded.\cite{Stamopoulos04SST,Stamopoulos05PRB} Referring to
homogenous layered hybrids recently spin-valve devices comprised
of FM-SC-FM TLs are studied
intensively.\cite{Gu02,Pena05,Potenza05}
In
Refs.\onlinecite{Gu02,Potenza05} the studied TLs employed Nb and
CuNi as the SC and FM constituents, while in
Ref.\onlinecite{Pena05} it was studied high-T$_c$ SC
YBa$_2$Cu$_3$O$_7$ in contact with FM
La$_{0.7}$Ca$_{0.3}$MnO$_{3}$. While in TLs comprised of low-T$_c$
Nb \cite{Gu02,Potenza05} the superconducting transition was
increased (decreased) when the FM layers where antiparallel
(parallel), in the TLs consisting of high-T$_c$
YBa$_2$Cu$_3$O$_7$\cite{Pena05} the opposite behaviour was
observed.

\begin{figure}[tbp] \centering%
\includegraphics[angle=0,width=7cm]{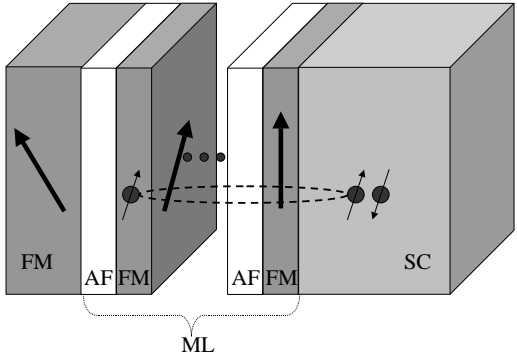}
\caption {Schematic representation of the ML-SC hybrid structure
studied in this work. Effective ferromagnetic or antiferromagnetic
coupling related to spin-triplet (parallel spins of the electrons)
or spin-singlet (antiparallel spins of the electrons)
superconductivity are also shown (see the discussion in the text
for details). Thick arrows refer to the magnetization of each FM
layer.}
\label{b4}%
\end{figure}

The present work offers experimental results on hybrids consisting
of MLs of manganites in direct contact with a layer of low-T$_c$
Nb SC. We chose MLs of manganites for the constructed hybrids
since we wanted to examine how the mechanism of exchange
biasing\cite{Schuller99,Moutis01} could influence the SC. In
addition, the layered structure of the ML offers enhanced
coercivity which is of practical importance for the design of
functional apparatus. In all samples a relatively thick FM buffer
layer has been used since we expected that this should act as a
main reservoir for generating stray fields that, in addition to
the exchange biasing mechanism, could influence the SC. The
specific manganites used for the ML are the insulating
antiferromagnetic (AF) La$_{0.33}$Ca$_{0.67}$MnO$_{3}$ and the
metallic FM La$_{0.60}$Ca$_{0.40}$MnO$_{3}$. Thus, each hybrid
consists of a FM buffer layer followed by a ML including $15$
bilayers
[La$_{0.33}$Ca$_{0.67}$MnO$_{3}$/La$_{0.60}$Ca$_{0.40}$MnO$_{3}]_{15}$
with a Nb layer placed on top (FM-ML-SC: which for simplicity is
noticed as ML-SC). The thicknesses used are $d_{fm}=50$
nm-[$d_{af}=4$ nm/$d_{fm}=4$ nm]$_{15}$-$d_{sc}=100$ nm. The
structure is schematically presented in Fig.\ref{b4}. Two
categories of ML-SC hybrids have been studied. The first category
has the insulating AF La$_{0.33}$Ca$_{0.67}$MnO$_{3}$ as the top
layer (noted as FM/AF-Nb), while the second one has the metallic
FM La$_{0.60}$Ca$_{0.40}$MnO$_{3}$ adjacent to the Nb layer (noted
as AF/FM-Nb). The same qualitative results were observed in both
categories of ML-SC hybrids. Simple FM-SC bilayers have also been
studied. These samples don't exhibit the effects observed in the
ML-SC structures. The MLs exhibit critical temperature
$T_c^{ML}=230$ K and were prepared by pulsed laser deposition on
LaAlO$_3$ $(001)$ substrates.\cite{Moutis01} The dc-sputtered Nb
films\cite{Stamopoulos04SST,Stamopoulos05PRB} have $T_c^{SC}=8.2$
K for the ML-SC samples and $T_c^{SC}=7$ K for the simple FM-SC
bilayers. A commercial SQUID (Quantum Design) was employed for the
magnetization measurements.

Our interesting magnetization data show that in the ML-SC hybrid
when the SC is cooled through its $T_{c}^{SC}$ under the presence
of a parallel magnetic field behaves as a {\it ferromagnet}. This
is feasible since the SC layer is ferromagnetically coupled to the
ML and as a consequence it has become insensitive to a reversal of
the external magnetic field until the coercive field of the ML is
exceeded. {\it When the ML's coercive field is exceeded the SC's
magnetization also reverses since it follows the magnetization of
the ML (switching effect)}. By performing experiments where the ML
was selectively biased or not we present evidence that the ML
structure separating the FM buffer and the SC layers imposes its
magnetic properties on the whole hybrid. Thus, the exchange
biasing mechanism could be used for regulating the magnetic
field's value where the switching of the SC's magnetization
occurs. The simple FM-SC bilayers studied in our work don't
exhibit these features. Thus, the ML structure is an important
ingredient for the generic observation of the switching effect.
Possible interpretations of the obtained results are discussed.

\begin{figure}[tbp] \centering%
\includegraphics[angle=0,width=9cm]{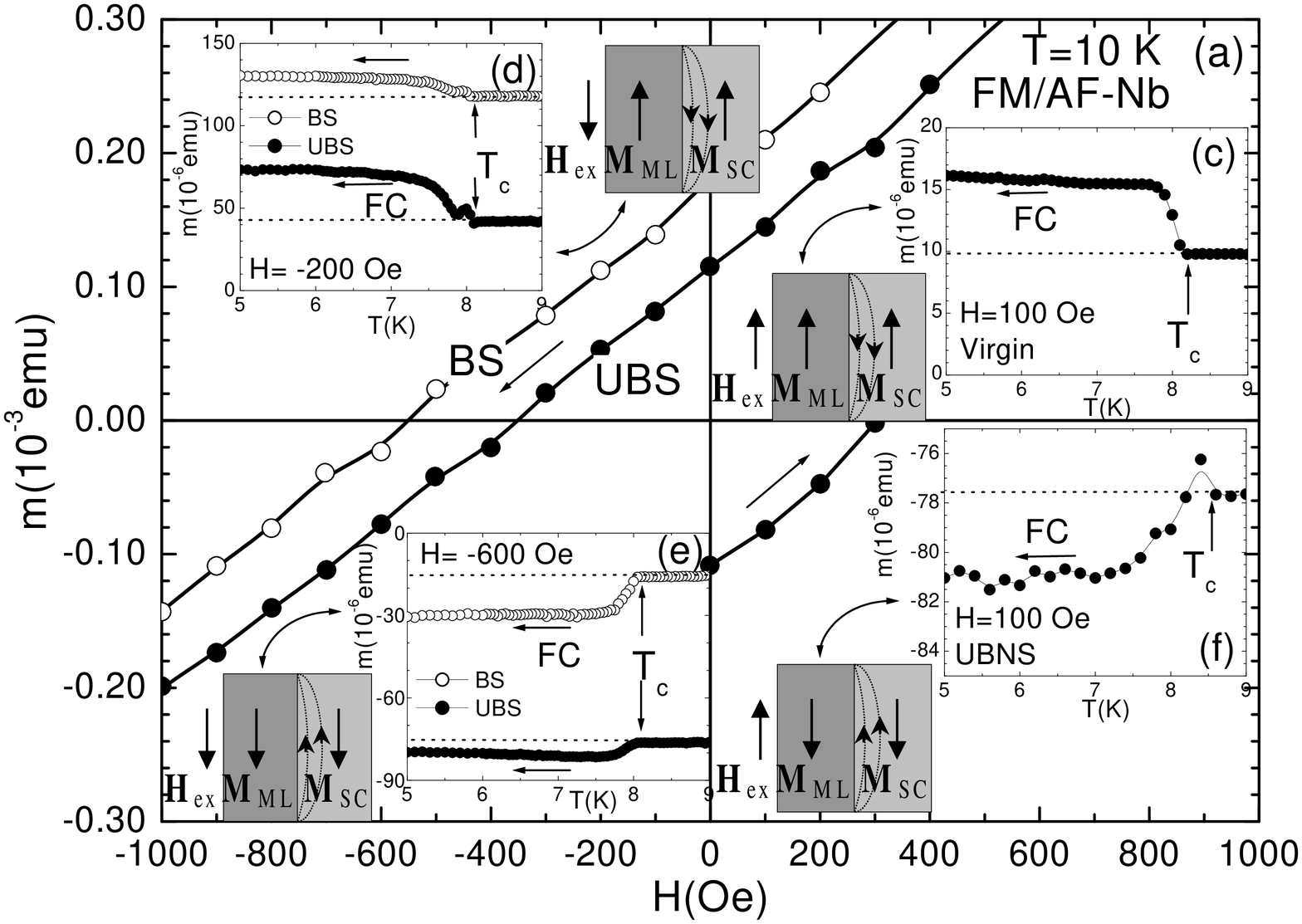}
\includegraphics[angle=0,width=8.9cm]{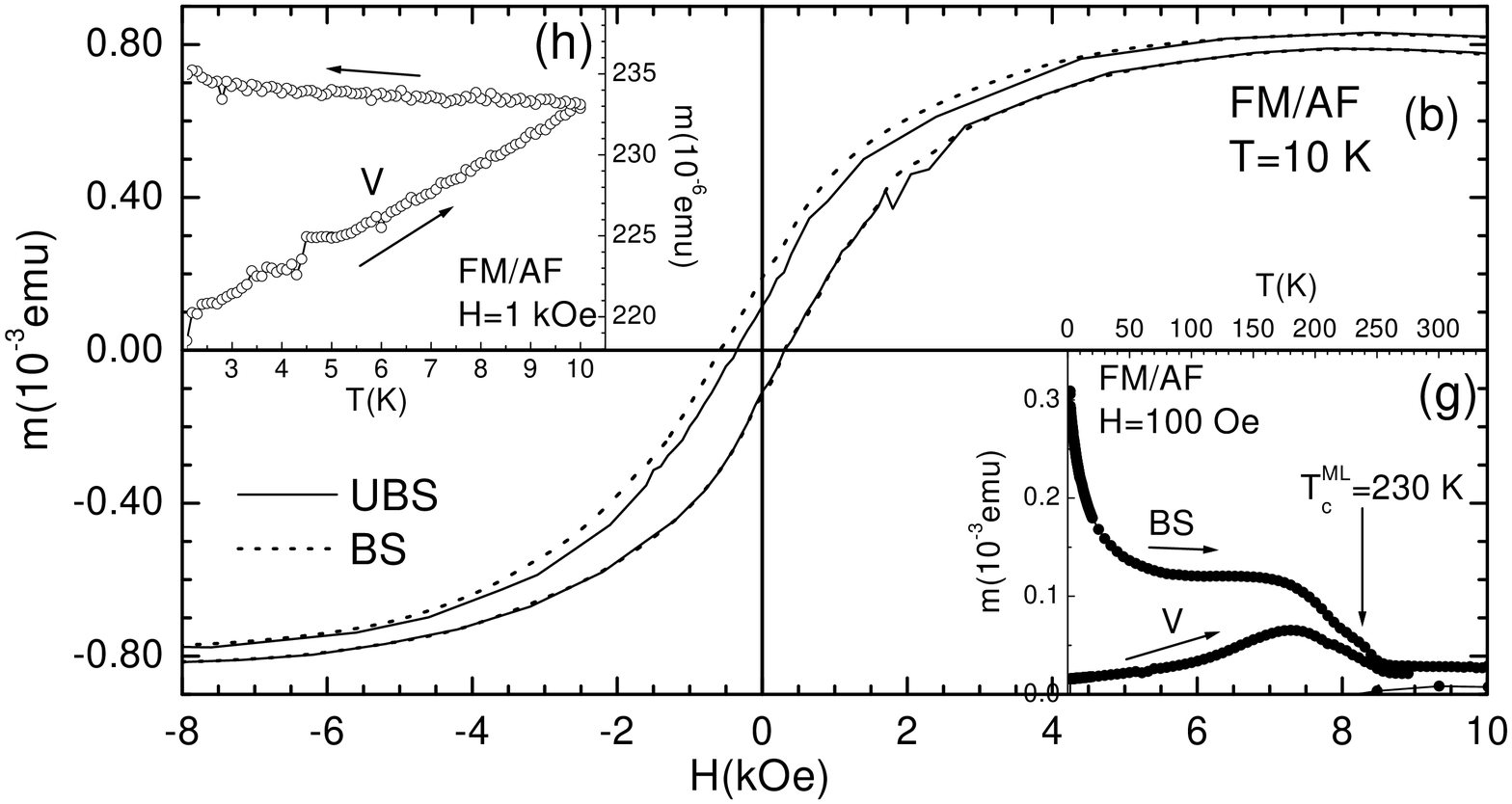}
\caption {Detailed magnetization data obtained in a FM/AF-Nb
hybrid. In the upper panel (a) we focus on the low-field regime of
the UBS and BS $m(H)$ branches obtained at $T=10$ K. The lower
panel (b) shows the respective data in an extended field range.
Insets (c)-(f) show $m(T)$ measurements: (c) for virgin (V) ML at
$H=100$ Oe, (d) for UBS and BS ML at $H=-200$ Oe, (e) for UBS and
BS ML at $H=-600$ Oe and (f) for $H=100$ Oe when the external
field is reversed once again. The relative configurations obtained
for $T<T_c^{SC}=8.2$ K of the ML's and SC's magnetizations in
respect to the external magnetic field $H_{ex}$ are shown in all
cases. Curved lines denote the stray fields of the ML that
penetrate the SC. Insets (g) and (h) show data for the pure ML
prior to the deposition of Nb. Inset (g) show the $m(T)$ BS and V
curves obtained at $H=100$ Oe for $5$ K$<T<340$ K, while inset (h)
presents a detailed m(T) curve obtained in the low-temperature
regime for $H=1$ kOe.
}
\label{b2}%
\end{figure}%

Figures \ref{b2}(a)-(h) introduces the main results of the present
work obtained in a FM/AF-Nb sample when the external magnetic
field $H_{ex}$ was parallel to its surface. In the upper panel (a)
shown are the low-field parts of the $m(H)$ branches where the
measurements have been performed for unbiased (UBS) and biased
(BS) conditions of the ML. In the UBS (BS) the hybrid was cooled
from above $T_c^{ML}=230$ K down to $T=10$ K in zero external
field, $H=0$ Oe ($H=50$ kOe). At $T=10>T_c^{SC}=8.2$ K the desired
magnetic field was set and the magnetization was recorded while
lowering the temperature until the transition of the SC was
accomplished. Thus, in all measurements the SC was field cooled
(FC). Insets (c)-(f) present detailed isofield $m(T)$ measurements
for different magnetic histories of the ML. The lower panel (b)
presents data for the pure ML prior to the deposition of the SC.
In the main panel we show the BS and UBS $m(H)$ loops in an
extended field regime. Inset (g) shows the respective $m(T)$ data
for the ML from $T=5$ K up to room temperature. These data reveal
that the mechanism of exchange bias is present in our ML. Inset
(h) shows a representative detailed m(T) curve in the
low-temperature regime obtained at $H=1$ kOe. Such detailed data
obtained in the pure ML are very important since after the
deposition of the SC we should be able to distinguish the origin
of any observed anomaly. We stress that we chose the FC protocol
for all the performed m(T) measurements for the following main
reasons: in this procedure the ML shows an almost constant
magnetization (see inset (h)). Thus any detected feature in the
m(T) curves could safely be ascribed to the magnetic behaviour of
the SC.

Let us start the discussion with inset (c). These data refer to
virgin ML and were obtained for positive orientation of the
external field $H=100$ Oe. We see that the SC's magnetization
presents a clear {\it increase} below its $T_c^{SC}$ despite the
fact that according to basic knowledge a {\it decrease} should be
observed due to the diamagnetic behaviour that the SC should
exhibit in respect to the external field. At first sight this
behaviour resembles the paramagnetic effect that is usually
observed in single Nb films and composite samples when FC through
their critical temperature.\cite{Thompson95,Schweigert00,Müller00}
As we show below the effects discussed in this work are completely
different. The respective measurements for negative orientation of
the external magnetic field are presented in insets (d) and (e)
for both BS and UBS initial conditions of the ML. The difference
between these sets of data is that in inset (d) where $H=-200$ Oe
the magnetization of the ML is still positive, while in inset (e)
where $H=-600$ Oe the coercive field has been exceeded so that the
magnetization of the ML has changed direction. In inset (d) we
clearly see again that below $T_c^{SC}$ the magnetization of the
SC presents an {\it increase}, while in inset (e) we see that
below $T_c^{SC}$ its magnetization {\it decreases}. Finally, inset
(f) presents the $m(T)$ curve obtained when the external field is
reversed again to positive orientation $H=100$ Oe. Since the
applied field is below the coercive field of the ML we observe
that the SC still exhibits a decrease below its $T_c^{SC}$. When
the external field exceeds the coercive field of the ML the SC
reverses its magnetization exhibiting an {\it increase} below its
$T_c^{SC}$.

\begin{figure}[tbp] \centering%
\includegraphics[angle=0,width=9cm]{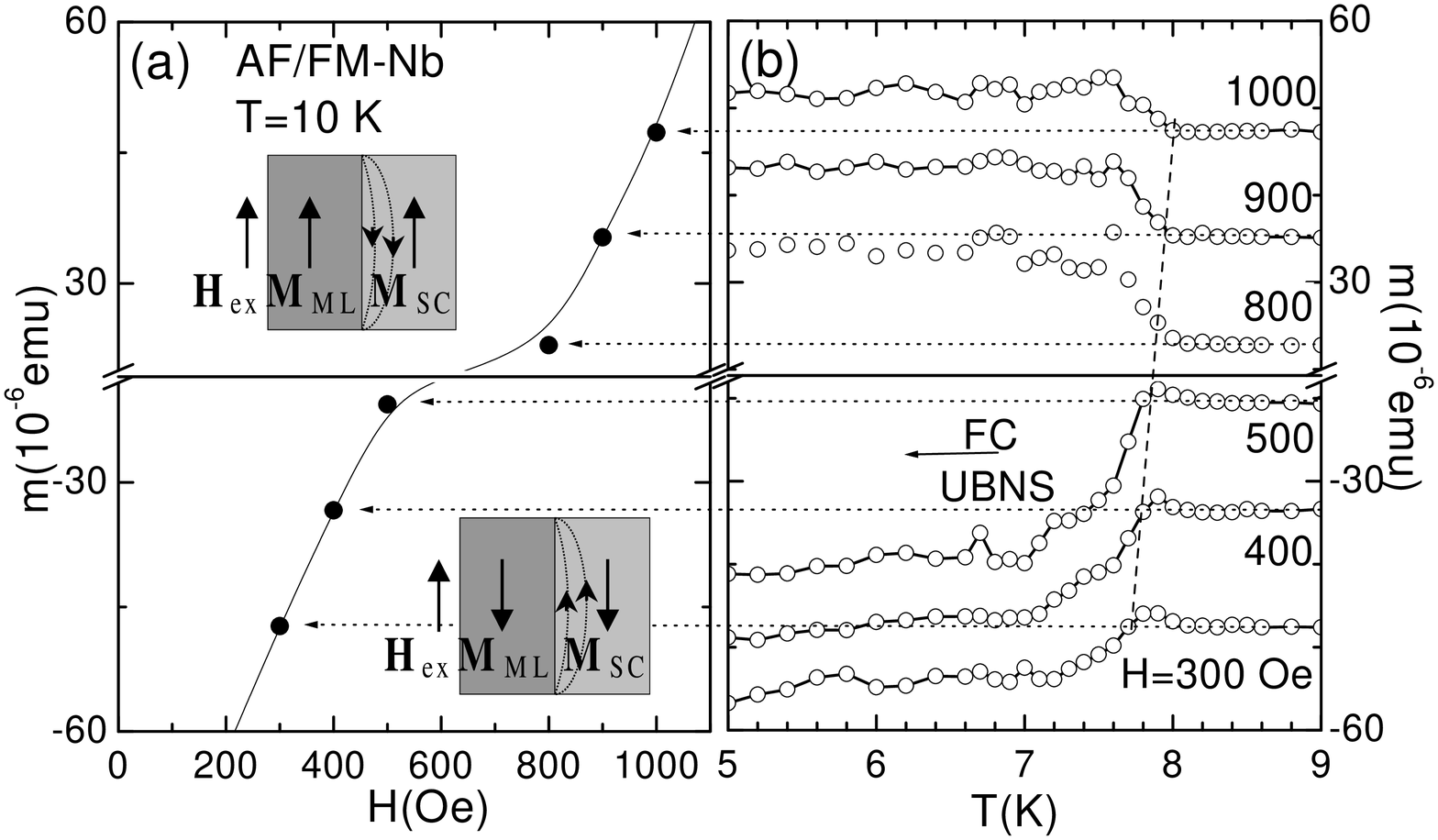}
\includegraphics[angle=0,width=9cm]{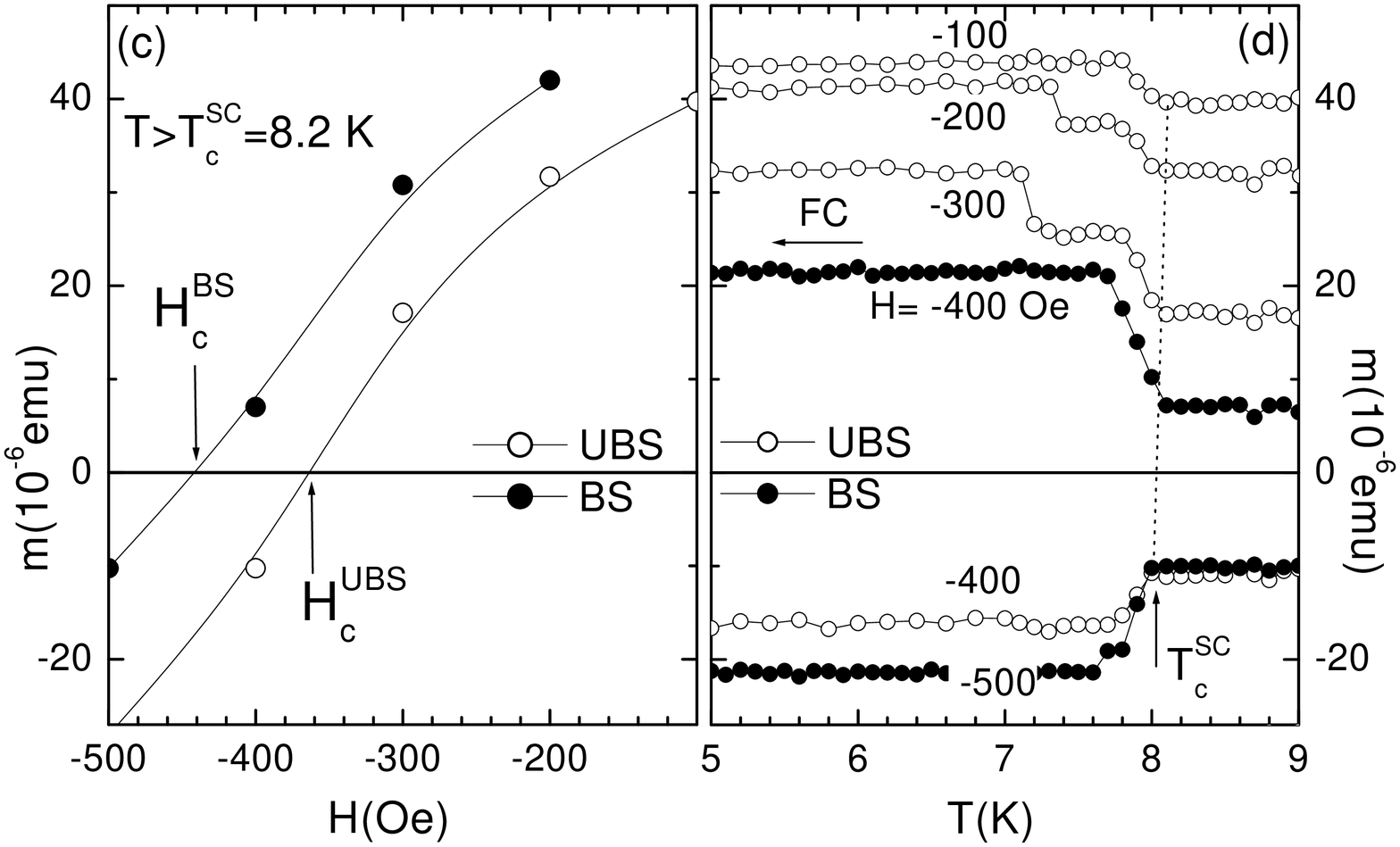}
\caption {Detailed data obtained in a AF/FM-Nb hybrid for various
magnetic fields in the regime of the ML's coercive field. Panel
(a) shows the increasing $m(H)$ branch while panel (b) shows the
respective $m(T)$ curves. The relative configurations of the ML's
and SC's magnetizations in respect to the external magnetic field
$H_{ex}$ are shown in both cases. Panels (c) and (d) present the
respective data for UBS and BS initial conditions.
}
\label{b3}%
\end{figure}

Detailed $m(T)$ data revealing the reversal of the SC's
magnetization near the coercive field of the ML are presented in
Figs.\ref{b3}(a)-(d) for an AF/FM-Nb sample. Panels (a) and (c)
show the increasing and decreasing $m(H)$ branches, respectively,
constructed from the $m(T=10$ K$)$ data, while panels (b) and (d)
show the respective $m(T)$ curves. In the data presented in panels
(a)-(b) the sample was initially UBS and was set to negative
saturation (the UBNS acronym in Fig.\ref{b3}(b) refers to initial
negative saturation). Then the field was lowered to the desired
value and the $m(T)$ curve was recorded while lowering the
temperature. These data clearly reveal that the SC's magnetization
follows that of the ML. Despite the fact that the external field
is positive the SC exhibits a negative step below its $T_c^{SC}$
as long as the ML's magnetization is also negative, i.e. below its
coercive field. When the ML's coercive field is exceeded the SC
also reverses its magnetization following the ML. The data
presented in (c)-(d) show a direct comparison for UBS and BS
conditions for the ML. These results reveal that when the ML is
BS, owing to the enhancement of its coercivity, the switching of
the SC's magnetization occurs at a higher magnetic field, $H=-450$
Oe when compared with the UBS state where $H=-350$ Oe. These
results propose that exchange biasing could be used as an
efficient control parameter for regulating the magnetic field's
value where the SC switches.

All these purely experimental data point to the situation which is
presented by the respective schemes of Figs.\ref{b2}(a) and
\ref{b3}(a). We see that in all cases the magnetization of the SC
is aligned parallel to the one of the ML as if the Nb layer is
simply an extra layer coupled ferromagnetically to the rest FM
layers through the adjacent AF ones. {\it Macroscopically, the
interpretation of the obtained results seems to relate to the
stray fields that the SC experiences.} The total stray fields of
both the FM buffer's and the ML's in each case are illustrated in
the schemes adjacent to insets (c)-(f) of Fig.\ref{b2}(a) by the
curved lines. Phenomenologically, it seems that the SC is forced
to behave diamagnetically not in respect to the external field but
in respect to the stray fields. {\it Microscopically, our results
may be related to recent theoretical works on the formation of
spin-triplet superconductivity in FM-SC hybrids when inhomogeneous
magnetization is offered to the SC by the
FM.}\cite{Volkov03,Bergeret01} In our ML-SC structures
inhomogeneous magnetization is experienced by the SC owing to the
specific modulated structure of the ML. The {\it ferromagnetic}
coupling observed in our data could be ascribed to the existence
of a spin-triplet component by employing an analogy of a model
that originally was proposed by theory in order the {\it
antiferromagnetic} coupling between a SC and a FM to be
explained.\cite{Bergeret04B} According to this model since the two
electrons forming a Cooper pair are well spatially separated
(especially close to T$_c^{SC}$) the first of them may be hosted
by the SC, while the second by one of the available FM
layers.\cite{Bergeret04B} Undoubtedly, the spin of the second
electron should be aligned ferromagnetically due to its
interaction with the magnetization of the host FM layer. As a
consequence the spin susceptibility of the first electron that
resides in the SC could exhibit {\it ferromagnetic} behaviour (as
it is observed in our results) only in case where the two
electrons of the pair have {\it parallel} spins i.e. in case of
triplet superconductivity. In the opposite case where spin-singlet
superconductivity prevails the spin of the first electron should
be {\it antiparallel} to the one's that is hosted in the FM. As a
result the SC should exhibit {\it antiferromagnetic} coupling to
the FM. The proposed scenario is presented qualitatively in
Fig.\ref{b4}. The antiferromagnetic behaviour was observed very
recently by J. Stahn et al. \cite{Stahn05} in neutron
reflectometry data obtained in
La$_{2/3}$Ca$_{1/3}$MnO$_3$/YBa$_2$Cu$_3$O$_7$ multilayers. Their
data indirectly suggested that possibly in those samples a
magnetic moment was created within the SC layers that was
antiparallel to the one of the FM layers as a consequence of the
conventional spin-singlet electron pairing.\cite{Stahn05} In our
La$_{0.60}$Ca$_{0.40}$MnO$_{3}$-[La$_{0.33}$Ca$_{0.67}$MnO$_{3}$/La$_{0.60}$Ca$_{0.40}$MnO$_{3}]_{15}$-Nb
hybrids by means of direct magnetization measurements we observed
a clear ferromagnetic coupling which, analogically, could be
compatible with the formation of a spin-triplet superconducting
component.\cite{Volkov03,Bergeret01,Bergeret04B}

\begin{figure}[tbp] \centering%
\includegraphics[angle=0,width=9cm]{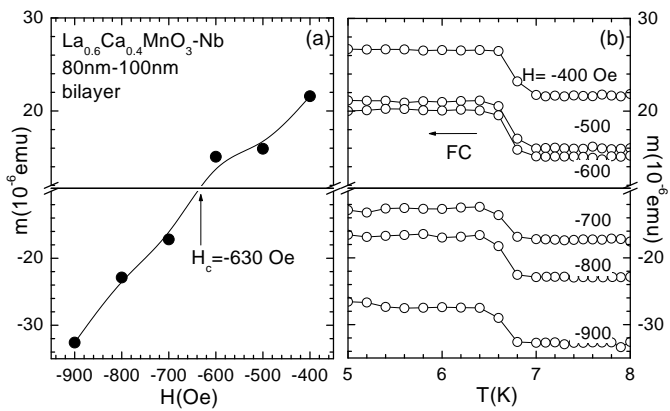}
\caption {(a) Decreasing branch of the m(H) curve and (b) detailed
FC m(T) data obtained near the coercive field of the FM for a
simple FM-SC bilayer. The switching of the SC is not observed in
such bilayers. 
}
\label{b1}%
\end{figure}%

As already discussed in the ML-SC structures studied in our work a
relatively thick La$_{0.60}$Ca$_{0.40}$MnO$_{3}$ FM layer was used
as a buffer since we expected that this should act as the main
reservoir for generating stray fields that could possibly
influence the SC. In addition, as a spacer between the FM buffer
and the SC top layers a ML was used since we wanted to incorporate
the exchange biasing mechanism. These complex ML structures are
needed in order the ferromagnetic coupling between the SC and the
FM to be observed. We note that the effect is {\it not} present in
simple FM-SC bilayers constructed of
La$_{0.60}$Ca$_{0.40}$MnO$_{3}$ and Nb ($d_{fm}=80$
nm-$d_{sc}=100$ nm). Representative data are shown in
Figs.\ref{b1} (a)-(b) near the coercive field of the FM where the
switching of the SC's magnetization should be observed. We clearly
see that the FM-SC bilayer don't exhibit the switching effect
(ferromagnetic coupling) but a conventional diamagnetic behaviour
in respect to the external magnetic field (which notice that in
this set of data is negative).

Concluding, we presented experimental results in hybrids
consisting of manganite MLs in contact with a Nb layer. In these
samples a thick FM layer has been used as a buffer. Our results
show that when the SC is FC it is ferromagnetically coupled to the
ML. The SC's magnetization is not actually affected when we
reverse the external magnetic field and only switches together
with the magnetization of the ML when its coercive field is
exceeded. The exchange biasing mechanism offered by the ML affects
the behaviour of the whole hybrid; the magnetic field's value
where the switching of the SC occurs may be regulated by
selectively biasing the ML or not. Macroscopically, our findings
may be explained by suggesting that the stray fields of both the
FM buffer's and the ML's may efficiently modulate the effective
magnetic field experienced by the SC. Microscopically, our results
may be explained by assuming the formation of spin-triplet
superconductivity. The ML structure is needed for the occurrence
of the discussed effects since simple FM-SC bilayers don't exhibit
the same behaviour. In addition to their importance to basic
physics our results are promising for the design of devices based
on magnetic switching processes.

\pagebreak

\end{document}